# Experimental Implementation of a Quantum Optical State Comparison Amplifier

*Ross J. Donaldson[1], Robert J. Collins[1], Electra Eleftheriadou[2], Stephen M. Barnett[3],*

*John Jeffers[2], and Gerald S. Buller[1]*

[1]*SUPA, Institute of Photonics and Quantum Sciences, David Brewster Building, Heriot-Watt University, Edinburgh, EH14 4AS, UK.*

[2]*SUPA, Department of Physics, John Anderson Building, University of Strathclyde, 107 Rottenrow, Glasgow, G4 0NG, UK.*

[3]*SUPA, School of Physics and Astronomy, University of Glasgow, Kevin Building, University Avenue, Glasgow, G12 8QQ, UK.*

**Quantum optical amplification that beats the noise addition limit for deterministic amplifiers has been realized experimentally using several different nondeterministic protocols[1–4]. These schemes either require single-photon sources[1–3], or operate by noise addition and photon subtraction[4]. Here we present an experimental demonstration of a protocol that allows nondeterministic amplification of known sets of coherent states with high gain and high fidelity[5]. The experimental system employs the two mature quantum optical technologies of state comparison[6] and photon subtraction[7] and does not rely on elaborate quantum resources such as single-photon sources. The use of coherent states rather than single photons allows for an increased rate of amplification and a less complex photon source[8]. Furthermore it means that the amplification is not restricted to low amplitude states. With respect to the two key parameters, fidelity and amplified state production rate, we demonstrate, without the use of quantum resources, significant improvements over previous experimental implementations.**

In classical electromagnetism a signal can be amplified, at least in principle, without being compromised by noise, allowing transmission losses to be overcome and signals to be propagated over greater distances. In quantum mechanics, many systems using quantum states to transmit a signal (for instance, quantum key distribution[9] or quantum digital

signatures[10,11]) would also benefit from amplification. Unfortunately perfect deterministic amplification of an unknown quantum optical signal is not possible[12]. Any attempt to amplify such a signal will introduce noise – the minimum amount of which is limited by the uncertainty principle[13]. This noise swamps any quantum properties that the signal might have.

An important idea was proposed, however, in 2009 by Ralph and Lund – the concept of nondeterministic amplifiers[14]. These work in postselection – the amplified output is only accepted based on the outcome of a measurement process. When the correct measurement outcome occurs the amplified output is accepted, otherwise it is discarded. The original scheme was based on the quantum scissors device[15]. Further protocols were proposed based on photon addition and subtraction[16] and on noise addition and photon subtraction[17]. All these have been later realized experimentally[1–4].

Each scheme has its advantages and drawbacks. The quantum scissors and photon addition based experiments require single-photon sources. The output is thus effectively limited to a set of states with almost no overlap with the two-photon state. Cascading the devices would circumvent this limitation, as would using quantum scissors with two photons as input[18]. Single-photon generation, however, is still an experimentally challenging proposition that typically only offers low photon fluxes[8] and the experimental success probability is very low, so cascading such devices is impractical. Research continues into the improvement of heralded photon source amplifiers aimed at measurement device independent quantum key distribution[19-21]. The noise addition scheme removes the requirement for single photons and works very well as a phase concentrator, but the fidelity of the output state compared to a perfectly-amplified version of the input state is typically low[4,5,22]. For example, for a coherent state with mean input photon number of 0.25 and an intensity gain of twofold the theoretical fidelity of the output to the target amplified state is approximately 0.8. As a comparison, the vacuum state has a fidelity with the same target state of more than 0.6.

In this paper, we demonstrate experimentally a recently-proposed protocol[5] that can amplify coherent states of any experimentally reasonable amplitude chosen from a limited set using non-demolition comparison[6] and photon subtraction[7], both established quantum optical techniques. The basic operation of the device is shown in Figure 1. The beamsplitter and detector $D_0$ perform the comparison between an input coherent state to be amplified and a selected guess state. The input state is chosen randomly from a known set. Guess states are chosen randomly from a set, chosen so that each interferes destructively with one of the possible input states so as to produce a vacuum state in the $D_0$ port. If the guess is correct the transmitted fraction of the input state interferes destructively with the reflected fraction of the guess state, photodetector $D_0$ does not fire (assuming no dark or background counts are present in the system) and the light is transmitted to the second beamsplitter. If the guess is incorrect some light leaks into the $D_0$ detector, where it may or may not cause a count, as low amplitude states have a large overlap with the vacuum. The non-firing of $D_0$ is taken as an *imperfect* indication that the guess state and the input state are matched.

The postselected output of the comparison beamsplitter is a reasonable approximate version of the amplified input state but inclusion of a second stage, comprising a highly-transmitting beamsplitter and a detector $D_1$ so as to perform photon subtraction, improves the fidelity. A small fraction of the incident light is reflected into $D_1$. When this detector fires it is more likely that the output of the first interferometer was of a higher mean photon number. This increases the purity of the output state, cleaning it of the lower mean photon number states produced by incorrect guesses at the comparison stage. The nominal gain of the whole device depends on parameters at both beamsplitters and is $g = t_2/r_1$.

The experimental system is shown in schematic form in Figure 2 and explained in more detail in the Methods. We generate input states in a comparatively simple manner by attenuating the output of a laser to the desired mean photon number per pulse. For coherent states this quantity is $|\alpha|^2$, where $\alpha$ is the coherent amplitude. These states are fed into a system comprised of two interwoven interferometers, the outer of which performs an analysis measurement on the amplified states. The lithium niobate (LiNbO$_3$) phase modulator sets the input state phase while the lower air-gap provides the phase for the guess state. In the current realization the states are interfered at a 50/50 beamsplitter, so the states in the input and guess state sets have the same mean photon number. The inner interferometer and detector then perform the state comparison. Photon subtraction is performed using a beamsplitter with a transmission to reflection ratio of 90:10, with the 10% reflecting to photodetector $D_1$. Generation of an amplified state is heralded by the absence of a detection event at detector $D_0$ and the presence of a detection event at detector $D_1$. Thus our implementation produces a device with a nominal photon number gain of $g^2$ equal to 1.8.

Post processing after this amplification allows for low noise measurement. The outer interferometer and photodetectors $D_A$ and $D_B$ perform analysis measurements on the output state. Events were examined conditional on events at photodetectors $D_A$ or $D_B$ occurring at periods when an event does not occur on detector $D_0$ and does occur on detector $D_1$.

The quality of the amplified output depends on the set of possible input states. We choose coherent states with mean amplitudes selected symmetrically on circles in the Argand plane. We present results for input sets $\{|\alpha\exp(2m\pi i/N)\rangle\}$, for $N=2$, $N=4$ and $N=8$, with $m=0\ldots N-1$. The results for the last of these two sets approach closely those of the phase-covariant set covering the entire circle.

In Figure 3a we show the outer analysis interferometer visibility for the two state set as a function of input mean photon number. We plot separately the visibilities for the output of the amplifier in three cases: when it is fully unconditioned, conditioned only on state comparison (only on detector $D_0$ not firing), and conditioned on both comparison and subtraction (on both detector $D_0$ not firing and detector $D_1$ firing for the same input optical pulse), so as to separate the effects on which the device is based. It is clear that the state comparison works better with increasing mean photon numbers, as expected. The photon subtraction step cleans the state further so that the visibility is close to ideal. This near perfect output is only possible with a single subtraction stage for $N = 2$ as the subtraction effectively excludes the wrong state reaching the outer interferometer.

Figure 3(b)-(d) shows the conditioned nominal output state fraction and the output fidelities for $N$=2, 4 and 8. The state fraction is the proportion of times that the device produces the desired amplified state $|g\alpha\rangle$. Without the conditioning imposed by the device, for the two-, four- and eight-state sets these percentages would be 50%, 25% and 12.5%, but it is clear that the amplification has increased these values to over 95%, 60% and around 30% respectively.

One effect not shown by the correct state fraction is that the probabilities of output states with amplitudes further from $|g\alpha\rangle$ are reduced relative to those nearer to $|g\alpha\rangle$. For the four and eight state sets this increases the fidelity without contributing to the correct state fraction. For each state set we provide an estimate of the fidelity of the output from the device compared to the nominal target output state. The estimate is obtained on the basis that the device can only produce a limited set of states, and on the measured counts in the outer interferometer. The calculations leading to this estimate are shown in the supplementary material.

The plots show that the fidelity and success rate of the amplifier system presented in this letter compare well with previous demonstrations of nondeterministic amplifiers. We emphasize here that the theoretical performance of the state comparison amplifier for the phase covariant input state set is similar to that for the four and eight state sets. For the four state set the conditioning increases the fidelity from an expected value for unconditioned output of 0.65, for a mean input photon number of around 0.5, to over 0.8. For the eight state set the unconditioned state ought to have a fidelity with the target state of 0.82 for a mean input photon number of 0.21. Again it is clear that the state comparison amplifier increases the value significantly, to over 0.9. For all three input state sets the fidelity is greatly increased across the whole range of photon numbers.

We can make some simple comparisons with two previous forms of amplifier for $g^2 = 1.8$. For $\alpha^2 = 0.5$ the scissors-based amplifier[1,18] produces an output state with a theoretical maximum fidelity with the target state of about 0.75[5], whereas the experimental figures for the state comparison amplifier are more than 0.98 for $N=2$ and more than 0.8 for $N=4$. For $\alpha^2 = 0.3$ the scissors-based amplifier[18] has a theoretical maximum fidelity of 0.9, and the figures for the state comparison amplifier are almost 0.99 and 0.9 for $N=2$ and 4 respectively. Comparison with the noise addition amplifier[4] can be simply made for $\alpha^2 = 0.25$, for which the state comparison amplifier produces an experimental fidelity of close to 0.9 for $N=4$, which is better than the theoretical value for the noise addition amplifier at this gain, although for higher gains the fidelity performance of these two devices ought to be similar.

The nominal success probability of the device is high (comparable to other nondeterministic amplification methods) and it depends on both the input mean photon number and the number of states in the set, but this is not the main advantage of the state comparison amplifier. Because we do not use quantum resources the high success probability translates into a high rate of success in real time (Fig. 4). High-quality transmission of quantum information at large data rates is possible. For example, for the two state set and a mean input photon number of 0.94, we obtain more than 26k conditioned counts per second in the $D_1$ detector when $D_0$ does not fire, corresponding to 26k almost perfectly amplified states. This rate can be increased straightforwardly by reducing losses or increasing the repetition rate of the pulses. The technological limitation in our system is simply the power damage threshold at the detectors. We can compare this rate to figures in other amplifier experiments. Systems that use downconverted photon pairs to produce single photons typically run at a relatively low rate of pair production. For example in reference 1 the rate of pair production was 2500 $s^{-1}$, and the success probability of the scheme together with detection losses mean success rates will be significantly lower than this. The systems in references 2 and 3 also use downconversion, and so their rates will be of the same order of magnitude. The success rate of the noise addition/photon subtraction experiments[4,23] is of the order of $1s^{-1}$.

Although the experiments described here were conducted using light with a wavelength of 850 nm, the experimental principles[5] are applicable to any wavelength of operation. The other main advantage of the system is that the rate of successful operation of the amplifier is the product of the amplifier success probability and the clock-rate of the driving laser – a feature that nondeterministic amplifiers based on the addition of single photons cannot replicate without the development of a rapid-fire synchronizable "photon machine gun". This is the fundamental reason why the success rate is so high for the state comparison amplifier.

The system could be extended and improved to operate at higher gains by using a comparison beamsplitter with a lower reflectivity, at some cost to the fidelity in particular for the larger state sets. The fidelity reduction can be offset by the inclusion of multiple photon subtraction stages, a technique that has been shown to be effective in other experiments[4].

The system has many possible applications, for example, it has the potential to be used in the sharing of quantum frames of reference[24]. Most strikingly perhaps, it could operate as a quantum repeater in quantum communications systems, such as quantum key distribution[9] or quantum digital signatures[10,11], and assist in increasing the transmission distance of such systems. A low-loss system similar to the one described here could be stationed approximately every few kilometres in standard optical fibre, acting as the perfect quantum equivalent of erbium doping.

**Acknowledgments**

This work was supported by the Royal Society, the Wolfson Foundation (UK) and the UK Engineering and Physical Sciences Research Council (EPSRC).

**Author Contributions**

The experimental system was conceived by J.J., based on previous work by E.E., S.M.B., and J.J. who also performed the theoretical analysis of the results presented in this work. Working under the supervision of G.S.B., R.J.C. wrote the control software and designed the experimental implementation which was assembled and operated by R.J.D.. R.J.C., R.J.D., and G.S.B. performed preliminary analysis of the raw results while processed results were analysed fully by E.E. and J.J.. All authors contributed the submitted manuscript.

**Methods**

The experimental system is shown in schematic form in Figure 2. The system comprised two interwoven interferometers; the inner insert interferometer (which effectively carried out the amplification stage) contained a phase modulator to establish the phase encoding while the outer (which measured the amplified states) had no active high frequency phase modulation. A temperature stabilized vertical cavity surface emitting laser (VCSEL) emitting at a wavelength of 850.38 nm, with spectral bandwidth 0.37 nm and ±0.8 pm central wavelength stability, launched the coherent states at a pulse repetition frequency (PRF) of 1 MHz through a free-space linear polarizer into 5 μm core diameter single-mode fiber (SMF). The linearly polarized light was coupled into 5 μm core diameter "panda eye" polarization maintaining fiber[25] (PMF) through an in-fiber polarizer so that the final polarization extinction ratio of the light in the PMF was in excess of 1,200:1 – contributing to a high mean interferometric fringe visibility of 92.41% in the inner interferometer and 92.24% in the outer (before conditional filtering). A computer controlled attenuator, based on a stepper motor controlled knife edge which spatially intercepted the beam, was used to set the mean photon number per pulse. The relative path lengths of the interferometers could be adjusted in approximately 15 nm steps over a 1.5 μm range by means of two computer controlled adjustable length air-gaps. These air-gaps also contained manual knife edge attenuators to balance the optical loss between the different fiber paths. A lithium niobate ($LiNbO_3$) phase modulator clocked at 1 MHz and phase locked to the laser provided the phase encoding of the states. The phase modulator required a voltage of 6 V to enact a phase change of π radians and had a variance in the desired phase of $\pm 1.6 \times 10^{-3}$ radians. The adjustable air-gaps in the interferometers were simultaneously adjusted under computer control for maximum visibility using a known sequence of phases before the phase to be amplified was transmitted. Photon subtraction was performed using an unequal ratio beamsplitter with 90% transmission to the outer interferometer and 10% reflection to photodetector $D_1$.

The photons were detected using commercially available free-running Geiger mode thick junction silicon single-photon avalanche diodes (Si-SPADs)[26] with a mean detection efficiency of 40.5% (at a wavelength of 850 nm) and a mean background count rate of 296 counts per second in the system. Although thin (or shallow) junction semiconductor diode photodetectors commonly offer superior timing jitter over their thick junction counterparts, they also typical exhibit lower detection efficiencies[8] and at the PRF used in these experiments the probability of intersymbol interference was negligible[27]. Semiconductor photodetectors were selected over other architectures due to the comparative ease of operation[8]. A wavelength of 850 nm was selected due to the relative immaturity semiconductor photodetectors for use in the traditional second and third telecommunications wavelength bands[28]. The PRF of 1 MHz was selected to ensure that the input power damage threshold of the photodetectors was not exceeded at the high mean photon numbers.

The arrival times of the photons at the detectors are recorded using a computer controlled time-stamping electronics[29] that could record with a time intervals of 1 ps. The maximum event rate that could be recorded by the combination of computer, custom software and time-stamping electronics was approximately 4 MHz, depending on other tasks undertaken by the computer's operating system while the software was in operation. The laser driver, phase modulator driver, and time stamping electronics all shared a common Rb reference clock to ensure that the electrical signals were phase locked.

After collection, the arrival times of the photons are software gated to discard those lying outside ±2 ns of the expected arrival time. The mean event retention of this gating process was 96.5% of events corresponding to incident coherent states. An analysis was performed with the laser disabled to measure using the effects of the gating process on non-time-correlated background events. The time gating process discarded 97% of the non-time-correlated background events.

To analyze the final amplified state the software carries out conditional filtering of the detector events, retaining only the events at photon detectors $D_A$ and $D_B$ that occur at periods when an event does not occur on detector $D_0$, and does occur on detector $D_1$. The window for conditional filtering was also ±2 ns. Electrical delays were used on the outputs of the photon detectors to ensure that the events for an individual laser pulse arrived simultaneously at the time-stamping electronics. The photon detectors exhibited some slight variation in the temporal response with varying photon flux[30] but the custom analysis software was able to take this into account during analysis. An analysis of six hundred individual one second duration measurements of non-time-correlated background events taken with the laser disabled found no occurrences of the pattern of $D_1$ firing while $D_0$ did not fire.

**Supplementary Material:**

**Fidelity Estimation From Interferometric Photocounts**

In order to calculate the fidelity of the device output with any nominal output state we need a way of estimating the state based on measurement results. We describe this here for $N = 2$. A similar method applies for other sets of states.

Firstly we note that "state" is not an observable, and therefore we cannot calculate the state based solely on measurement results. We need an initial assumption. Ours is that the output of the state comparison amplifier is one of two possible states, either the amplified output state $|g\alpha\rangle$ or the vacuum state $|0\rangle$, and that these are *a priori* equally likely.

Our measurement mixes the output state with another copy of $|g\alpha\rangle$, the test state, at a 50/50 beamsplitter such that all of the light should exit at (say) detector $D_A$. We can calculate the probabilities of various detector counts at $D_A$ and $D_B$ based on this assumption. These are

$$\begin{aligned}
P(1,0\,|\,g\alpha) &= 1 - \exp(-2\eta l\, g^2\, \alpha^2) - \varepsilon, \\
P(0,1\,|\,g\alpha) &= \varepsilon \exp(-2\eta l\, g^2\, \alpha^2), \\
P(1,1\,|\,g\alpha) &= \varepsilon \left[1 - \exp(-2\eta l\, g^2\, \alpha^2)\right], \\
P(1,0\,|\,0) &= P(0,1\,|\,0) = \exp(-\eta l\, g^2\, \alpha^2/2)\left[1 - \exp(-\eta l\, g^2\, \alpha^2/2)\right], \\
P(1,1\,|\,0) &= \left[1 - \exp(-\eta l\, g^2\, \alpha^2/2)\right]^2
\end{aligned} \quad (S1)$$

where, for example, $P(1,0\,|\,g\alpha)$ is the probability that $D_A$ received a photocount and $D_B$ did not given that the output of the state comparison amplifier was the state $|g\alpha\rangle$. Here we assume for simplicity that the detectors have the same quantum efficiency and that there is a loss factor $l$ before the detectors. The quantity $\varepsilon$ takes account of the possibility that either the state comparison amplifier output or the test state might not be precisely $|g\alpha\rangle$ - in other words it represents the imperfection in the interferometer that performs the measurement. We can estimate this directly from measurement results if we wish. However, it turns out not to be necessary for our fidelity calculations.

Consider only the set of results when the state comparison amplifier output was the state $|g\alpha\rangle$, and suppose that this set of results yields $n_A^{g\alpha}$ counts at $D_A$ and $n_B^{g\alpha}$ counts at $D_B$. We do not know the total number of pulses $N_{g\alpha}$ that contributed to these counts, but it is related to the numbers of counts via

$$n_B^{g\alpha} = \left[ P(0,1|g\alpha) + P(1,1|g\alpha) \right] N_{g\alpha} = \varepsilon N_{g\alpha}$$
$$n_A^{g\alpha} = \left[ P(1,0|g\alpha) + P(1,1|g\alpha) \right] N_{g\alpha} = \left[ 1 - (1+\varepsilon)\exp(-2\eta l g^2 \alpha^2) \right] N_{g\alpha}. \quad (S2)$$

These can be solved to give

$$N_{g\alpha} = \frac{n_A^{g\alpha} + n_B^{g\alpha} \exp(-2\eta l g^2 \alpha^2)}{1 - \exp(-2\eta l g^2 \alpha^2)}. \quad (S3)$$

Thus the number of $|g\alpha\rangle$ output pulses $N_{g\alpha}$ can be determined from experimental results. Similarly if we consider only the set of results when the state comparison amplifier output was the state $|0\rangle$ and suppose that this yields $n_A^0$ counts at $D_A$ and $n_B^0$ counts at $D_B$ we can find $N_0$, the number of pulses that contributed to these counts:

$n_A^0 = n_B^0 = \left[ 1 - \exp(-2\eta l g^2 \alpha^2) \right] N_0$, so taking the average we obtain

$$N_0 = \frac{n_A^0 + n_B^0}{2\left(1 - \exp(-2\eta l g^2 \alpha^2)\right)}. \quad (S4)$$

We can now compute the density operator as a simple ratio of pulse numbers in the form

$$\hat{\rho} = P(g\alpha)|g\alpha\rangle\langle g\alpha| + P(0)|0\rangle\langle 0|, \tag{S5}$$

where the probabilities are

$$P(g\alpha) = \frac{N_{g\alpha}}{N_{g\alpha} + N_0}$$
$$P(0) = \frac{N_0}{N_{g\alpha} + N_0}. \tag{S6}$$

The fidelity is then straightforwardly calculated as

$$\begin{aligned} F &= \langle g\alpha|\hat{\rho}|g\alpha\rangle \\ &= \frac{N_{g\alpha}}{N_{g\alpha} + N_0} + \exp(-2g^2\alpha^2)\frac{N_0}{N_{g\alpha} + N_0}. \end{aligned} \tag{S7}$$

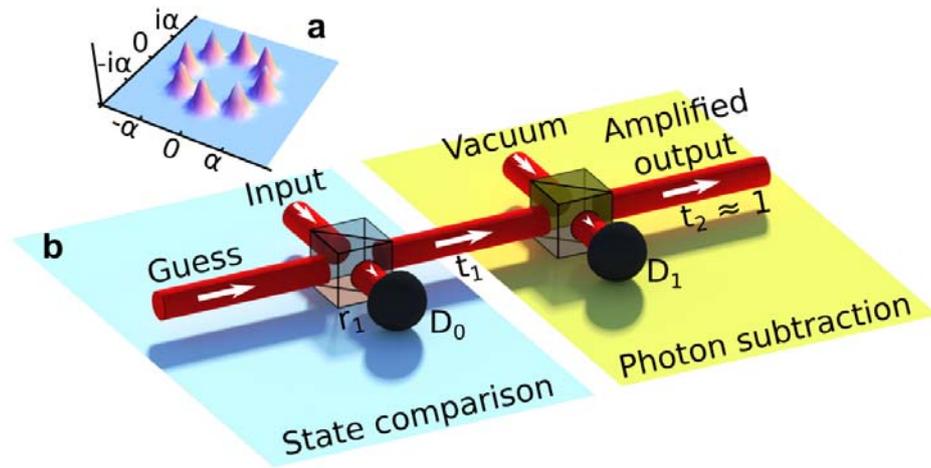

**Figure 1: Schematic of the state comparison amplifier and eight possible input states. (a) Eight possible input coherent states (b) Input and guess coherent states are combined at a beamsplitter.** If the guess is correct the vacuum exits the $D_0$ port and all light passes to the subtraction stage. When $D_0$ does not fire and $D_1$ does the output is passed, otherwise it is rejected. The combination of comparison and subtraction cleans the output of incorrect guesses.

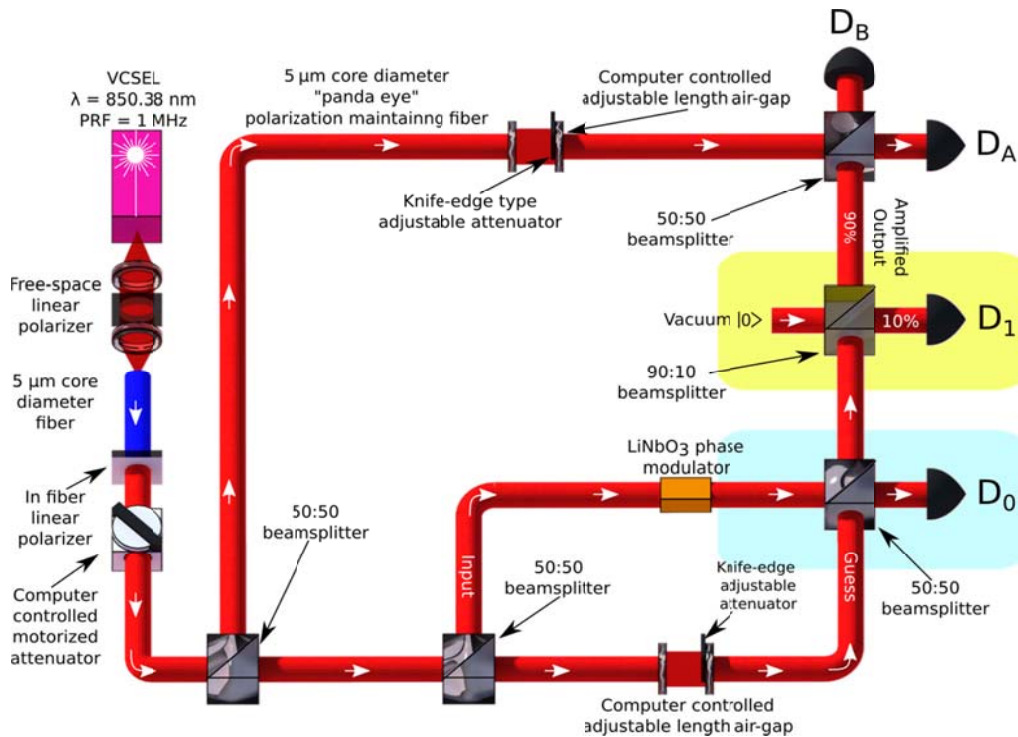

**Figure 2| Experimental implementation of the quantum optical state comparison amplifier and output state analyzer.** The system comprises two interferometers, one with a lithium niobate (LiNb0$_3$) phase modulator inside another with only a path-length stabilizing air-gap. D$_0$, D$_1$, D$_A$, and D$_B$ are free-running Geiger mode, thick junction silicon single-photon detectors. VCSEL denotes a vertical cavity surface emitting laser, PRF the pulse repetition frequency of the laser, and $|0\rangle$ the vacuum state.

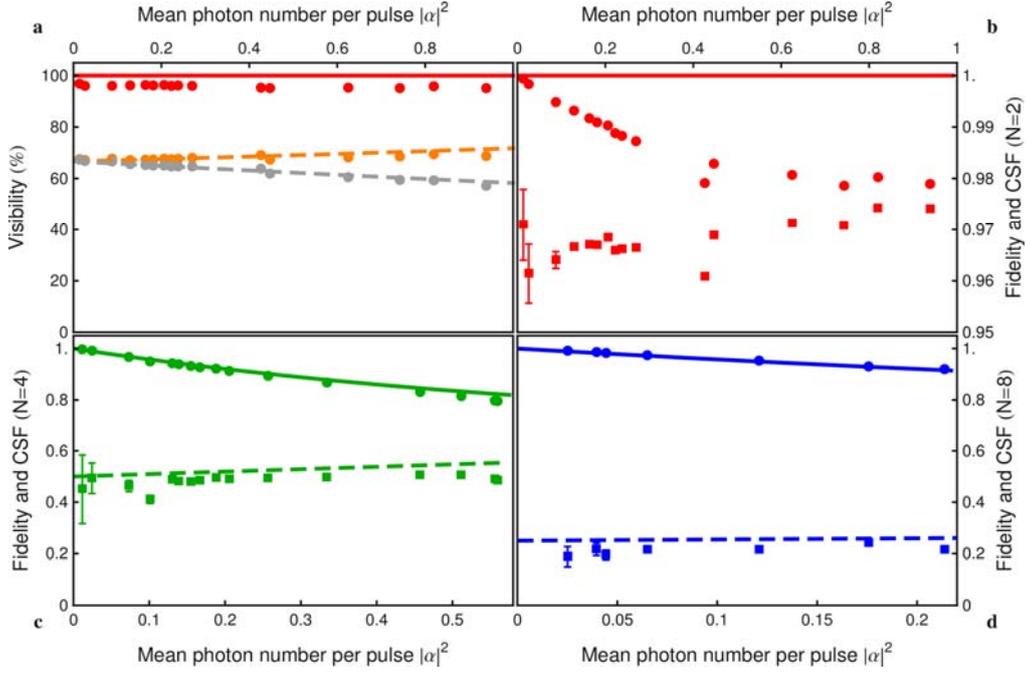

**Figure 3| Experimental results from the quantum optical state comparison amplifier and output state analyzer**

**(a) Visibilities for the two state set.** Here we show the experimental and fitted theoretical visibilities at the outer interferometer for $N=2$. The grey dotted curve is for the unconditioned output, the orange dashed curve is for output conditioned on Detector $D_0$ not firing, and the red curve is for $D_0$ not firing and $D_1$ firing. Typical standard error in the measurements is ±0.05. Lines are theoretical best-fit curves based on experimental parameters.

**(b-d) Fraction of the correct state in the output and Fidelity.** The three figures represent the fractions of the correct state in the output of the state comparison amplifier (squares) and the fidelity of the output state to the target state $|g\alpha\rangle$ (dots) as a function of input photon number. Frame (b) corresponds to the two-state set, frame (c) - four and frame (d) - eight. The $|\alpha|^2 = 0.0033$ point for the eight state set has been omitted as the low number of overall counts renders this unreliable. The standard error, shown for the correct state fraction, decays quickly with mean photon number. Standard errors for the fidelity are small: typically

±0.0003 for $N=2$, ±0.0022 for $N=4$ and ±0.0013 for $N=8$. Lines are theoretical best-fit curves based on experimental parameters.

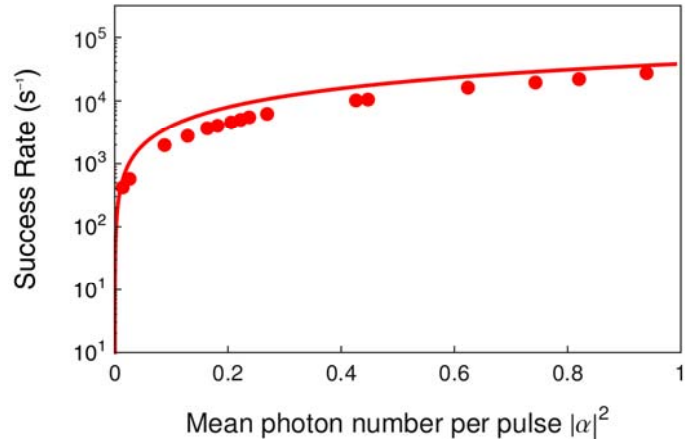

**Figure 4 Amplifier Success Rate.** The success rate of the amplifier corresponds to the gated count rate at the $D_1$ detector when $D_0$ does not fire, and is shown for the two state set as a function of input photon number. The success rates for the four and eight state sets are almost identical to that for *N*=2.